\documentstyle[preprint,aps,epsfig]{revtex}
\begin{document}
 
\title{Antikaon flow in heavy-ion collisions: effects of absorption
and mean-field potential\footnote{Dedicated to Gerry Brown on the
occasion of his 70th birthday.}}
\bigskip
\author{G. Q. Li and C. M. Ko}
\address{Cyclotron Institute and Physics Department,\\
Texas A\&M University, College Station, Texas 77843, USA}
 
\maketitle
 
\begin{abstract}
We study antikaon flow in heavy-ion collisions at SIS energies
based on the relativistic transport model (RVUU 1.0).
The production of antikaons from both baryon-baryon and
pion-baryon collisions are included. Taking into account only
elastic and inelastic collisions of the antikaon with nucleons
and neglecting its mean-field potential as in the cascade model,
a strong antiflow or anti-correlation of antikaons with
respect to nucleons is seen as a result of
the strong absorption of antikaons by nucleons.
However, the antiflow of antikaons disappears after
including also their propagation in the
attractive mean-field potential.
The experimental measurement of antikaon flow in heavy-ion
collision will be very useful in shedding lights on the relative
importance of antikaon absorption versus its mean-field potential.
\end{abstract}
 
 
The collective flow of particles in heavy-ion collisions
\cite{gut89,eos1,eos2,fopi,ags,dan86} has been proven to be
a useful observable for studying both the nuclear equation of
state at high densities and hadron properties in dense matter.
Detailed analyses of proton flow in heavy-ion collisions
using transport models have already shown that the experimental data
are consistent with a soft nuclear equation of state if
the momentum-dependence is properly included in the
nuclear mean-field potential \cite{pan,zhang}.
 
The flow of produced particles have also been extensively studied
in transport models. For particles that are strongly
absorbed by nucleons such as the pion, antikaon, and antiproton,
studies based on the cascade model, in which their mean-field potentials
are neglected, predict an appreciable
antiflow or anti-correlation of these particles with respect to nucleons
as a result of strong absorption by the spectator nucleons
\cite{bass,liba,dan95,rqmd,arc}.
In particular, antikaon flow was studies in Ref. \cite{rqmd}
for Au+Au collisions at AGS energies using the relativistic
quantum molecular dynamics (RQMD) and was found
to be anti-correlated with the nucleons.
A similar calculation was carried out in Ref. \cite{arc}
using A Relativistic Cascade (ARC) model. Again, the antikaon
flow was found to be opposite to that of nucleons. 
 
On the other hand, kaons
and lambda hyperons, which can not be absorbed by nucleons, are
seen in cascade calculations
to flow in the same direction as nucleons
\cite{eos2,arc}. However, after including propagation
in mean-field potentials, their flow patterns become
very different \cite{li95a,brown96a,li96}. Because of a weak repulsive
potential resulting from the cancellation between an attractive
scalar and a repulsive vector potential,
kaons are `pushed' away from nucleons, leading to 
the disappearance of kaon flow in heavy-ion collisions, which has
recently been confirmed by experimental data from
Ni+Ni collisions at 1.93 GeV/nucleon measured by
the FOPI collaboration at GSI \cite{fopi}.
On the other hand, the potential for a lambda in nuclear matter is
known to be attractive so they are attracted towards nucleons, leading
to a lambda flow almost as strong as that of nucleons \cite{li96}.
This also seems to agree with the preliminary data from
both the EOS \cite{eos2} and the FOPI \cite{fopi} collaboration.
 
The consistent and simultaneous explanation of the FOPI data
on kaon and lambda flow in Ni+Ni collisions at 1.93 GeV/nucleon
provide a strong evidence for the role of mean-field potentials
in heavy-ion collisions. This is complimentary to that provided by
studies for the particle yields and spectra \cite{fang94,li94a,li94b,li95b}.
Therefore, it is important to see if the predicted
antiflow of pions, antikaons, and antiprotons based
on the cascade-type treatment is also affected by
mean-field potentials.

For pion flow in heavy-ion collisions
at SIS energies, the medium effect has recently been studied
in Ref. \cite{fae}
by introducing in the quantum molecular dynamics (QMD)
an in-medium pion dispersion
relation based on either the delta-hole model \cite{weise} or a
phenomenological model \cite{gale}.
The resulting attractive pion potential
reduces the effect from pion absorption and
thus changes significantly the pion flow pattern.
The resultant pion flow could
be in the same direction as nucleons if the attractive pion potential
is sufficiently strong.

Recently, Brown and Rho \cite{brown96b} have shown via chiral Lagrangian
with dropping pion decay constant in medium that the
attractive scalar (before being cut down by the range term) and repulsive
vector potentials acting on a kaon are just 1/3 of
nucleon mean-field potentials in the Walecka model.
For an antikaon, the vector potential becomes also attractive
due to G-parity. It is thus interesting to see if the resulting
attractive antikaon mean-field potential has any effects on
antikaon flow in heavy-ion collisions.
In this Rapid Communication, we shall report the results from
such a study.

To study medium effects in heavy-ion collisions at SIS energies,
the relativistic transport model (RVUU 1.0) \cite{ko}
has been extensively used by us in the past
\cite{li95a,fang94,li94a,li94b,li95b}.
This model includes explicitly the nucleon, delta resonance,
and pion. Furthermore, it can treat
eta, kaon, antikaon, hyperon, antiproton, and dilepton
production in heavy-ion collisions at SIS energies using
the perturbative test particle method. The model is based on
the non-linear $\sigma$-$\omega$ model, but extended to include
medium effects on the delta
resonance, pion, and strange particles.

Medium modifications on kaon and antikaon are obtained from the
mean-field approximation to the chiral Lagrangian, including
the Brown and Rho scaling for the pion decay constant
\cite{brown96a,brown96b,brown91}.
In this approximation, the antikaon dispersion relation in nuclear
medium is modified by both attractive scalar and vector potentials,
proportional, respectively, to the nuclear scalar $\rho_S$ and
baryon $\rho_B$ densities \cite{li95a}, i.e.,
\begin{eqnarray}\label{dis}
\omega ({\bf k},\rho _B) = \Big[ m_K^2+{\bf k}^2 -{\Sigma _{KN}
\over f_\pi^{*2}} \rho _S + \big({3\rho _B\over 8f_\pi^{*2}}\big)^2\Big]
^{1/2} -{3\rho_B\over 8f_\pi^{*2}},
\end{eqnarray}
where $\Sigma _{KN}\approx 450$ MeV is the kaon-nucleon sigma term
\cite{dong,fuku},
and the in-medium pion decay constant $f_\pi^*$ is related to that 
in free space $f_\pi \approx 93$ MeV by
$f_\pi^{*2}/f_\pi^2\approx 0.6$ \cite{brown96a,brown96b}.
This scaling relation is derived from the Gell-Mann$-$Oakes$-$Renner
relation and the Feynman-Hellmann theorem, and is
supported by recent transport model analyses \cite{li95c}
of the CERES \cite{ceres} and HELIOS-3 \cite{helios} dilepton data.
In Eq. (\ref{dis}), we have neglected
corrections to the scalar attraction from the range term as it is small
for antikaon at high densities \cite{brown96a}.
 
As in Ref. \cite{li95a}, we define the antikaon potential as
\begin{eqnarray}\label{pot}
U({\bf k}, \rho_B)=\omega ({\bf k},\rho_B) - \omega _0({\bf k}),
\end{eqnarray}
where $\omega _0({\bf k})= (m_K^2+{\bf k}^2)^{1/2}$. 
At normal nuclear matter density $\rho _0$, the potential is about -190
MeV for a $K^-$ at rest. This is very close to the -200$\pm$20
MeV found in Ref. \cite{gal} from the kaonic atom data.
 
In addition to antikaon production from baryon-baryon  
collisions as considered in Ref. \cite{li94a}
using the cross section parameterized in Ref. \cite{zwer},
we also include in this study
antikaon production from pion-baryon collisions.
Experimental data are available for $\pi^- p
\rightarrow p K^0 K^-$, $\pi^- p \rightarrow n K^+ K^-$,
$\pi^- p \rightarrow n K^0 {\bar K}^0$, and
$\pi^+ p \rightarrow p K^+ {\bar K}^0$ \cite{data}.
Within experimental uncertainties, these data are more or less
consistent with the charge-independent assumption and each can be 
reasonably fitted by the following parameterization
\begin{eqnarray}\label{xsection}
\Sigma _{\pi ^- p \rightarrow n K^+K^-} =
{0.08 (\sqrt s -\sqrt {s_0})^2 \over 0.043 + (\sqrt s- \sqrt {s_0})^2}
~~ {\rm mb},
\end{eqnarray}
where $\sqrt s$ and $\sqrt {s_0}=m_N+2m_K$  are in units of GeV.
The isospin-averaged cross section is then given by
\begin{eqnarray}\label{isospin}
\sigma _{\pi N\rightarrow NK{\bar K}} = {7\over 3}
\sigma _{\pi ^- p \rightarrow n K^+K^-}.
\end{eqnarray}
The isospin-averaged cross section for $\pi \Delta \rightarrow 
NK{\bar K}$ is assumed to be the same as that for $\pi N\rightarrow N
K{\bar K}$. We find that for Ni+Ni collisions at 1.93 GeV/nucleon,
the contributions to antikaon production from baryon-baryon and 
pion-baryon collisions are about 70\% and 30\%, respectively.
As far as antikaon flow is concerned, the uncertainties in the 
elementary antikaon production cross sections do not play 
a significant role. However, it will be important in the future to
carry out detailed analyses of these cross sections if one is
interested in the absolute yield of antikaons from heavy-ion collisions.
 
We consider two scenarios for antikaon production and propagation in
heavy-ion collisions. In the first scenario, we neglect possible 
medium modifications of the antikaon properties both in calculating
its production and in treating its propagation in nuclear
medium. In this cascade-type calculation, antikaons only undergo
elastic and inelastic (mainly strange-exchange processes) collisions
with baryons. The cross sections for the latter processes
such as ${\bar K}N\rightarrow {\bar K}N$,
${\bar K}N\rightarrow \Lambda \pi$, and ${\bar K}N\rightarrow
\Sigma \pi$ are taken from the parametrizations given in
Ref. \cite{cug90} which fit the experimental data quite well.
 
In the second scenario, we include the antikaon scalar and vector potentials
in determining the threshold for its production.
For antikaon production cross sections in baryon-baryon collisions,
which are parameterized in terms of the maximum antikaon momentum $p_{max}$,
we evaluate its value using in-medium masses as in Ref. \cite{li94a}.
For antikaon production from meson-baryon interactions,
we again use in-medium masses to
calculate the threshold energy $\sqrt {s_0}$ in Eq. (\ref{xsection}).
This lowers the antikaon
production threshold as the antikaon potential is attractive,
leading to an enhanced production of antikaons
as already demonstrated in Ref. \cite{li94a}.
In addition, we also include antikaon propagation in the mean-field
potential to modify their momentum distribution. The equations
of motion for an antikaon in nuclear medium are given by \cite{li95a}
\begin{eqnarray}\label{eom}
{d{\bf r}\over dt} = {{\bf k}\over E^*}, ~~
{d{\bf k}\over dt} = -\nabla _x U ({\bf k}, \rho _B),
\end{eqnarray}
where $E^*=\Big[m_K^2+{\bf k}^2 -{\Sigma _{KN}
\over f_\pi^{*2}} \rho _S + \big({3\rho _B\over 8f_\pi^{*2}}\big)^2\Big]
^{1/2}$.
 
In-plane flow is usually presented by the average transverse momentum
$\langle p_x \rangle$ as a function of rapidity $y_{cm}$
in the center-of-mass frame of the colliding nuclei.
As a reference, we first show in Fig. 1a the nucleon flow in
Ni+Ni collisions at 1.93 GeV/nucleon and impact parameter $b=4$ fm.
The results for antikaon flow are shown in Fig. 1b by the dashed
and the sold curve for the scenario without and with
antikaon mean-field potential, respectively. In the
cascade-type calculation, the antikaon flow is found to be in 
the opposite direction to that of nucleon, as was found in Refs.
\cite{rqmd,arc} for heavy-ion collisions at AGS energies. The 
flow parameter $F$, defined as the slope parameter at mid-rapidity, is
found to be about -50 MeV. The appearance of antiflow of antikaons in
heavy-ion collision in cascade calculation is due to the strong
absorption of antikaons in the direction of nucleon flow. 
 
Including antikaon propagation in the mean-field potential, the
picture changes dramatically. Because of an attractive potential,
antikaons are `pulled' towards nucleons, as was found for the
lambda hyperon \cite{li96} and pion \cite{fae}, both feeling
attractive potentials in nuclear medium. The effect from antikaon
propagation
in mean-field potential is seen to be stronger than the effect
from its absorption by nucleons, and
the final antikaon flow turns out to
be quite weak with a flow parameter of about 15 MeV.  
 
The effects of absorption and mean-field potential on antikaons can also
be studied from their azimuthal distribution.
As shown in Ref. \cite{li96b}, this effect can be most
clearly seen near the target and projectile rapidities, due to the
large anisotropy in the nucleon azimuthal distribution.
For comparisons, we show in Fig. 2a the nucleon azimuthal distribution
$dN/d\phi$ near the target rapidity. A significant excess of nucleons
is seen in the lower hemisphere near the target.
As shown in Ref. \cite{li96b}, a similar excess of
nucleons appears also in the upper hemisphere near the projectile rapidity.
On the other hand, the nucleon distribution is
almost isotropic near mid-rapidity.
 
The antikaon azimuthal distribution
near the target rapidity is also shown in Fig. 2.
Fig. 2b gives the distribution of primordial antikaons that
do not suffer any final-state interactions. The distribution is more or less
isotropic, as was also found in Ref. \cite{rqmd} for primordial
antiprotons. The distribution of antikaons
including the absorption effect but not the mean-field effect is shown
in Fig. 2c. The dip near $\phi =180^0$ shows that there is a strong
anti-correlation of antikaons with nucleons as a result
of the absorption of antikaons by
the spectator nucleons, which are located in the lower hemisphere.
This has also been observed in Ref. \cite{rqmd} for
antiprotons in heavy-ion collisions at AGS energies. 
Fig. 2d shows the azimuthal distribution of
antikaons including both absorption and mean-field effects. As in the
case of in-plane flow, the effect from the mean-field potential
is stronger than that from absorption, and the final antikaon
azimuthal distribution turns out to be almost isotropic.
Although not shown here, similar effects on antikaon azimuthal
distribution are again seen near the project rapidity.

In summary, using the relativistic transport model (RVUU 1.0),
we have studied both the antikaon in-plane and out-of-plane flow in
heavy-ion collisions at SIS energies.  In particular
the effects of antikaon absorption and mean-field potential 
on the flow pattern are investigated. In the cascade-type treatment, we
observe clearly the anti-correlation of antikaons with respect
to nucleon, as was seen in both RQMD and ARC calculation \cite{rqmd,arc}.
This is mainly due to the strong absorption of antikaons by
nucleons. If we further include the propagation of antikaons
in the attractive mean-field potential, the strong anti-correlation
between antikaons and protons seen in both in- and out-of-plane flow
disappears, and the antikaon distribution becomes almost
isotropic. Since there is little doubt that antikaons are strongly
absorbed by nucleon, the experimental observation of a
disappearance of antikaon flow will provide a
strong evidence for the presence of antikaon
mean-field potential in medium.
Similar considerations can be applied to antiprotons, and such
a study will be reported elsewhere.

\vskip 1cm
 
We are grateful to Gerry Brown for his continuous support and
enlightening discussions. The support of C.M.K. by a Humboldt 
Research Award is also gratefully acknowledged, and he thanks 
Ulrich Mosel of the University of Giessen for the warm hospitality 
during his visit. This work was supported in part by the 
National Science Foundation under Grant No. PHY-9509266.

\newpage
\begin{figure}
\epsfig{file=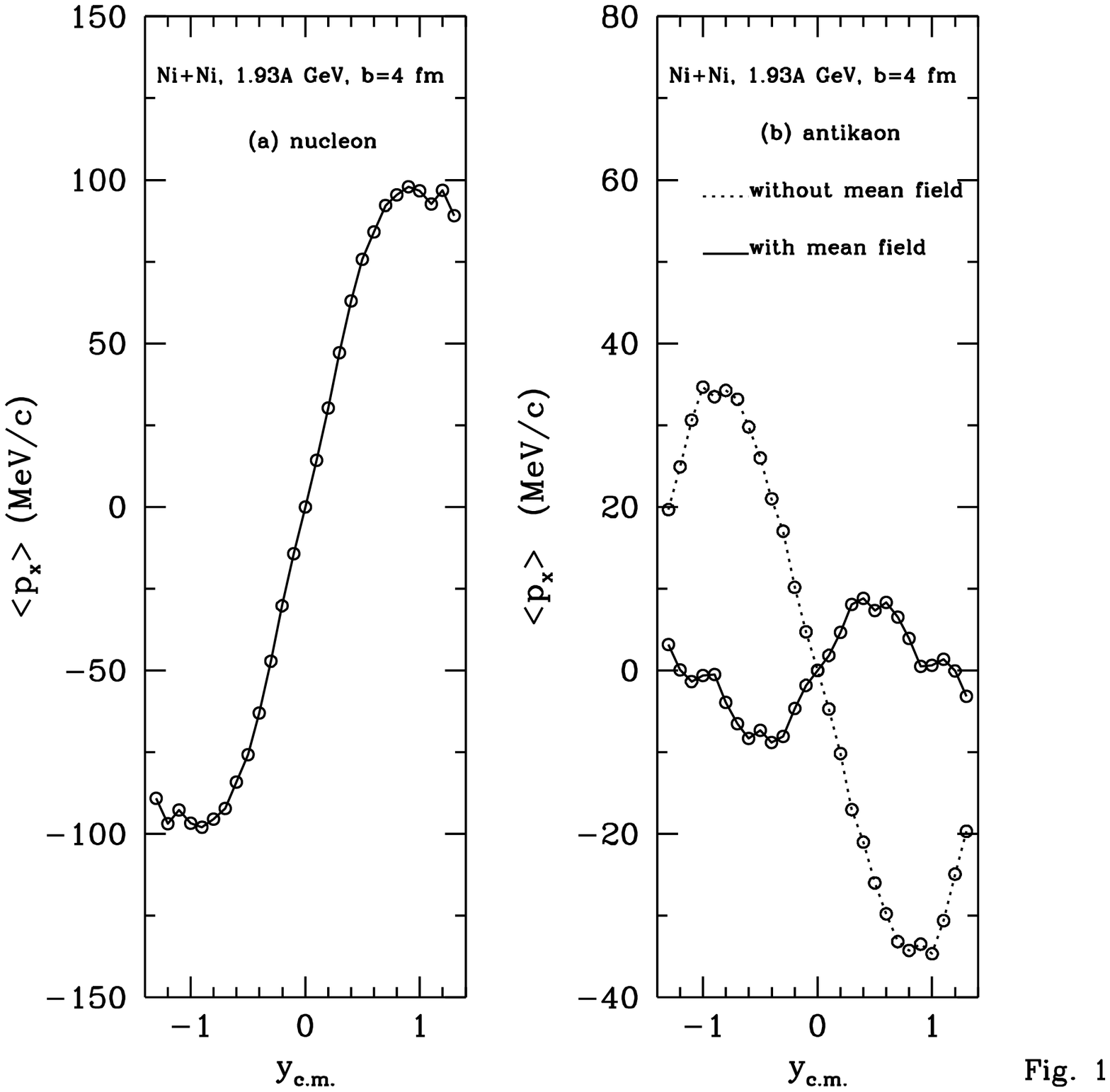,height=4in,width=6in}
\vskip 0.5cm
\caption{Nucleon (a) and antikaon (b) flow in Ni+Ni collisions
at 1.93 GeV/nucleon and impact parameter b=4 fm.}
\end{figure}

\newpage 
\begin{figure}
\epsfig{file=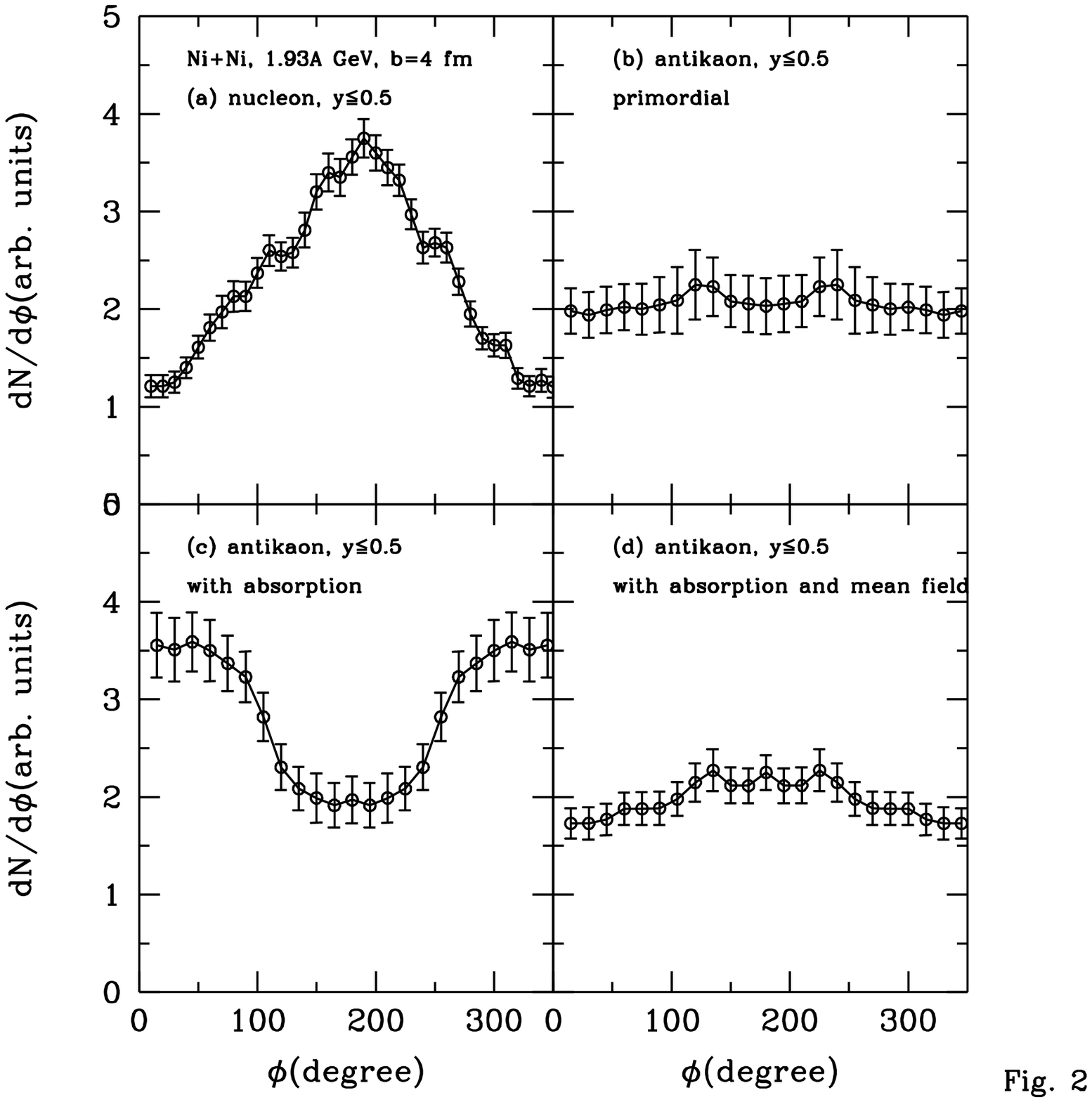,height=5in,width=6in}
\vskip 0.5cm
\caption{Nucleon (a) and antikaon (b-d)
azimuthal distributions near target rapidity
for the reaction in Fig. 1.}
\end{figure}

\end{document}